\documentclass[
    ,final            
  ]
  {aipproc}

\layoutstyle{8x11single}

\begin{document}

\title{Forward Jet-like Event Spin-dependent Properties in Polarized p+p Collisions at $\sqrt{s}\,$=200 GeV}


\author{Nikola Poljak for the STAR collaboration}{
  address={University of Zagreb, Croatia}
}

\begin{abstract}
The STAR collaboration has reported precision measurements on the transverse single spin asymmetries for the production of forward $\pi^0$ mesons from polarized proton collisions at the center of mass energy of 200 GeV. Disentangling contributions to forward asymmetries requires one to look beyond inclusive neutral pion production to the production of forward jets or direct photons. During the RHIC running in the year 2006, STAR with the Forward Pion Detector++ (FPD++) in place collected 6.8 inverse picobarn of data with an average polarization of 60\%. FPD++ had sufficient acceptance for ``jet-like'' objects, which are clustered responses of an electromagnetic calorimeter primarily sensitive to incident photons, electrons and positrons. For the ``jet-like'' objects, we reconstructed the angle of the outgoing leading neutral pion with respect to the fragmenting parton, thus enabling us to disentangle the contributions to the forward $\pi^0$ asymmetries. The simulated data set shows that on average there are approximately 2.5 fragmenting mesons per one ``jet-like'' object. Preliminary FPD++ results provide no evidence of measured contributions to the asymmetry from jet fragmentation, implying the Sivers distribution functions play a substantial role in producing the large inclusive $\pi^0$ asymmetries previously measured by STAR.
\end{abstract}

\maketitle

\section{Introduction and experimental setup}
For a particle produced in a collision of transversely polarized protons, the analyzing power $A_N$ is equal to the difference of spin-up and spin-down cross sections divided by their sum. $A_N$, just one example of a transverse single spin asymmetry, is expected to be near zero in a leading-twist collinear perturbative QCD description of particle production \cite{Kane78}. The measured cross sections for production of neutral pions ($\pi^0$) produced with large Feynman-x (2$p_L /\sqrt{s}$) and moderate $p_T$ in p+p collisions are found to be in agreement with next-to-leading order pQCD calculations at $\sqrt{s}=200\,$GeV and are included in global fits on fragmentation functions \cite{Adams06, deFlorian07}. ``Jet-like" structures are observed in two-particle correlations involving a forward pion, as expected from pQCD. Precision measurements of the $\pi^0$ asymmetry as a function of $x_F$ and $p_T$ were reported, showing large $A_N$ at large $x_F$ \cite{Abelev08}. The measured $x_F$ dependence matches the Sivers effect \cite{Sivers90, Sivers91} expectations qualitatively. Theoretical understanding of the pion production data continues to evolve \cite{Kang11}.

Presently, two classes of models try to explain the observed asymmetries. The Sivers effect manifests itself as an asymmetry in the forward jet or gamma production while the Collins effect \cite{Collins93} manifests itself as an asymmetry in the forward jet fragmentation. Both effects introduce a transverse scale $k_T$ required to produce the observed asymmetries. The Collins mechanism introduces this scale in the fragmentation functions, making it sensitive to correlations of the hadron transverse momenta and parton transversity. To distinguish between the mechanisms, one has to look beyond inclusive $\pi$ events to direct gamma or ``jet-like'' objects. Since the Collins effect is a spin-dependent azimuthal modulation of hadrons around the thrust axis of an outgoing quark, integrating over the azimuthal angle would cancel it, leaving only the Sivers effect. This is one possible path to distinguishing the two effects. ``Jet-like" events are the clustered response of an electromagnetic calorimeter which is primarily sensitive to incident photons, electrons and positrons. The forward electromagnetic detector present during the RHIC running in the year 2006 was specifically designed to have sufficient acceptance for ``jet-like'' objects. Theoretical predictions for the estimated $\pi^0$ Collins contribution to the asymmetry for the observed process show it to be negligible \cite{dAlesio11}. The data obtained with the detector were used to separate the Sivers/Collins contributions to the asymmetries observed for the $\pi^0$ events. 

The forward STAR calorimeters prior to 2006 measured the inclusive $\pi^0$ cross section as well as the single beam spin asymmetry for their inclusive production. In 2006 the detectors placed on the West STAR platform were upgraded to a detector called the Forward Pion Detector ++. A discussion of the detector setup and calibration is given in \cite{Poljak11}.

To have a better understanding of the unpolarized result obtained in the data, full PYTHIA/GEANT simulations have been made with adequate statistics. We used PYTHIA 6.222, which predates tunings related to ``underlying event'' for midrapidity Tevatron data, since these tunings impact forward production at RHIC energies. The newer versions of PYTHIA reduce agreement between data and simulation in the forward region.

\section{Results}

FPD++ was a modular detector that consisted of two modules placed symmetrically with respect to the beam line at a distance of 7.4 m from the interaction point. ``Jet-like'' clusters are formed in an event by considering energy depositions $>$ 0.4 GeV in all cells of the FPD++. A cluster consists of N cells, where N is the maximal subset of cells found to be within a cone of radius 0.5 in $\eta - \phi$ space. The $x_F$ and $p_T$ for the cluster are given by the vector sum of momenta from each cell, assuming the energy deposition is from photons originating from the collision vertex. Clusters with at least 10 cells having cluster $p_T >$ 1.5 GeV/c and $x_F >$ 0.23 are required in the analysis. A further requirement that the cluster centroid is within the calorimeter volume by at least two large cell widths is also imposed. 

The energy deposition profile and the invariant mass distributions of the ``jet-like'' objects were looked at in both data and the simulations. For the simulations, the energy deposition computed by GEANT for events generated by PYTHIA was digitized and the pseudodata was run through the exact same algorithms as the data. It was demonstrated that throughout the $x_F$ range of the event sample the properties of the ``jet-like'' objects in the data and the simulations compared well. A subset of data that contained both a reconstructed $\pi^0$ and a ``jet-like" object in a single event was extracted. These events were further characterized with the help of PYTHIA, showing that on average, the $\pi^0$ carries $\approx 90$\% of the energy of the ``jet-like'' object (Fig.1). With the help of association analysis, we have shown that the reconstructed ``jet-like'' thrust axis agrees well with the direction of either a hard-scattered parton or a radiated parton and that on average, there are 2.5 fragmenting meson per one ``jet-like'' object, making them reasonably jetty.  

The component of the $\pi^0$ momentum perpendicular to the ``jet-like'' object axis ($k_T$) was found to be in the domain of transverse momentum dependent fragmentation and has also shown reasonable agreement in data and simulations (Fig.1). A systematic study of the ``jet-like'' object model was done to ensure that no special point in the model parameter space was selected. The study was made by changing the model parameters by 10\% both on data and simulations. It was demonstrated that the changes in the unpolarized results are small and smooth. Further, the data and simulations followed the same trends when changing a single parameter. This demonstrated that the result isn't strongly dependent on the model parameters.

\vskip 2mm
\begin{figure}[h!]
\hskip 8mm
\begin{minipage}{19pc}
\includegraphics[width=16.5pc]{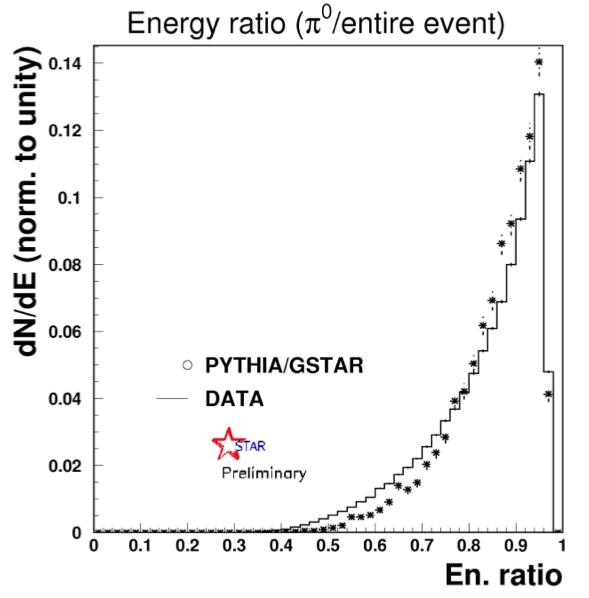}
\end{minipage}\hspace{2mm}
\begin{minipage}{19pc}
\includegraphics[width=16.5pc]{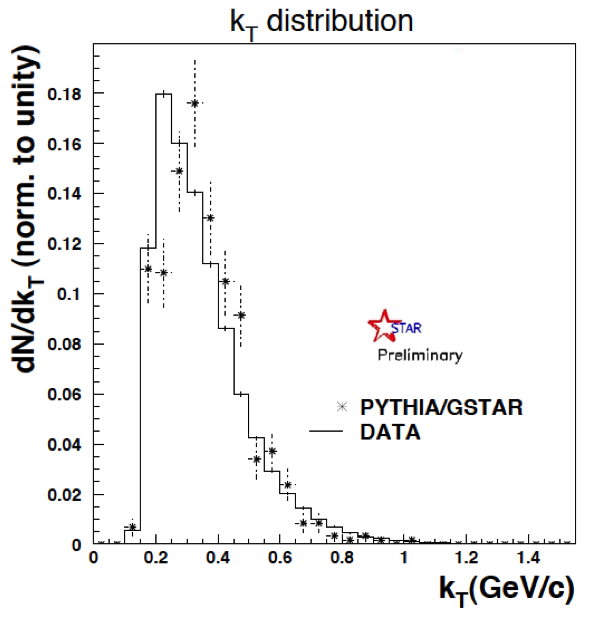}
\end{minipage} 
\caption{Left panel: The energy ratio distribution of the events entering the analysis. The distribution shows what fraction of the event enegy is carried by the leading particle. Right panel: The component of the $\pi^0$ momentum perpendicular to the ``jet-like'' object axis ($k_T$). In both panels the error bars are statistical.}
\end{figure}

To be able to separate the Collins/Sivers contributions to the measured $\pi^0$ asymmetries, we defined the angle $\gamma$ as the azimuthal angle of the $\pi^0$ with respect to the reaction plane (Fig.2). $\gamma$ is defined mirror symmetrically (CW for the left module, CCW for the right module), so that $\gamma \approx 0$ corresponds to $\pi^0$ having larger $p_T$ than the ``jet-like'' object. Further, for a given bin in $\gamma$ defined this way, an event reconstructed in the left module from an up-polarized proton measures the same fragmentation as an event reconstructed in the right module from a down-polarized proton. By forming the geometric mean from these events in each of the $\gamma$ bins in the asymmetry calculation, the detector effects are minimized. 

The unpolarized $\gamma$ angle distribution was found both in data and simulations and shows reasonable agreement \cite{Poljak11}. The peaking near $\gamma =0$ in the shape of the distribution is a detector acceptance effect. In most cases both the $\pi^0$ and the thrust axis of the ``jet-like'' object are reconstructed in the small cells. The $\chi^2$ for fitting a constant to the spin-averaged $\gamma$ distribution systematically decreases in an analysis that restricts the acceptance to have the thrust axis of the ``jet-like'' object increasingly centered on a calorimeter module. Consequently, the shape of the spin-averaged $\gamma$ distribution is due to finite detector acceptance. The $p_T$ dependence of the ``jet-like'' production cross section strongly favors low $p_T$ events.  This dependence, coupled with detector acceptance, enhances the number of events with $\gamma \approx 0$. 

\vskip 1mm
\begin{figure}[h!]
\hskip 8mm
\begin{minipage}{19pc}
\includegraphics[width=18.5pc]{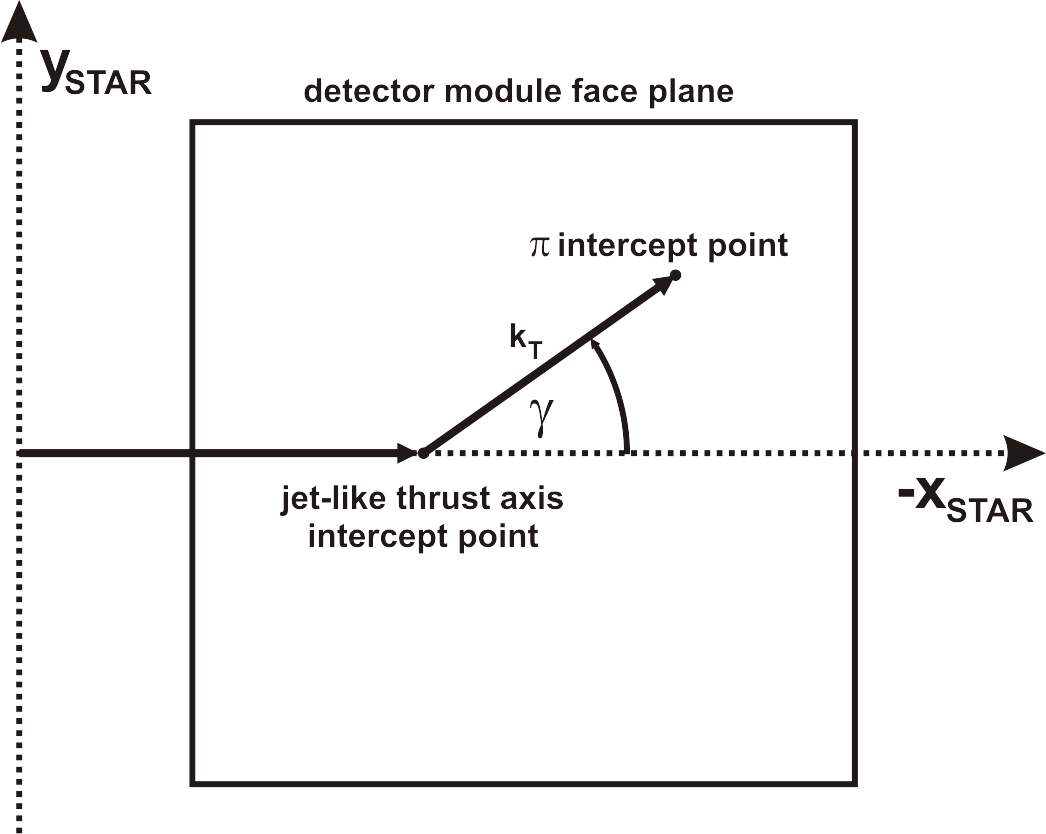}
\end{minipage}\hspace{1mm}
\begin{minipage}{19pc}
\includegraphics[width=17.5pc]{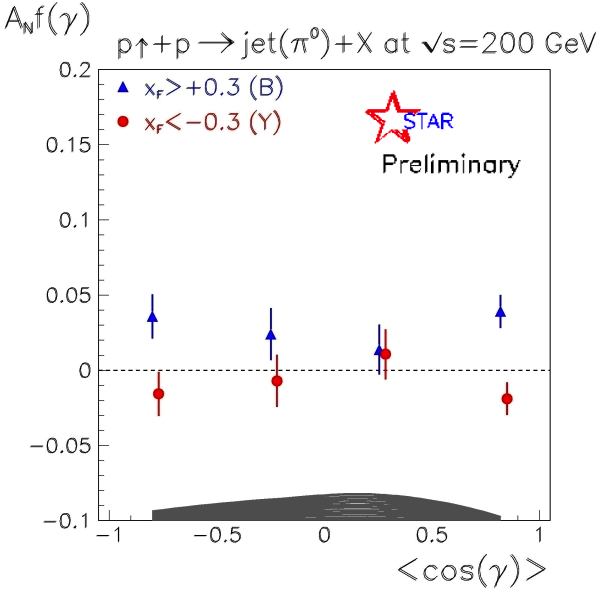}
\end{minipage} 
\caption{Left panel: The schematic view of the $\gamma$ angle in the right detector module.  Right panel: The dependence of the asymmetry on the $\gamma$ angle. The lines on the points represent statistical errors and the black area the systematic errors.}
\end{figure}

The dependence of the asymmetry on the $\gamma$ angle was found. The data was separated in 4 equally sized bins in $\cos(\gamma)$. On the plot of $A_N$ vs. $\cos(\gamma)$ (Fig.2) the Collins contribution would be proportional to the slope parameter, since the fragmentation would be angle-dependent. If the slope is consistent with zero and the asymmetry is larger than zero, one has isolated the Sivers effect, since the fragmentation is not involved, and one is left with just the initial state $k_T$. The dependence of the calculated asymmetry on $\cos(\gamma)$ is given on the figure. The lines on the points represent statistical errors and the black area the systematic errors. The $x_F>0$ and the $x_F<0$ points have been slightly horizontally shifted for improved visibility. The asymmetry for negative $x_F$ values is consistent with zero, while the positive $x_F$ values show a positive asymmetry, but no dependence on $\cos(\gamma)$. Hence, the Collins effect is not found to be present in the production of forward neutral pions. 


\bibliographystyle{aipproc}   

\bibliography{sample}

\IfFileExists{\jobname.bbl}{}
 {\typeout{}
  \typeout{******************************************}
  \typeout{** Please run "bibtex \jobname" to optain}
  \typeout{** the bibliography and then re-run LaTeX}
  \typeout{** twice to fix the references!}
  \typeout{******************************************}
  \typeout{}
 }

\end{document}